\documentclass[12pt,peerreview,draftcls,onecolumn,journal]{IEEEtran}
\ifCLASSOPTIONcompsoc
\else
\fi
\ifCLASSINFOpdf
\else
\fi

\usepackage[caption=false,font=footnotesize]{subfig}
\usepackage[dvips]{graphicx}
\DeclareGraphicsExtensions{.eps}
\usepackage{float}
\usepackage{amsmath}
\usepackage{multicol}
\usepackage{multirow}
\usepackage{color}
\usepackage{hyperref}
\begin{document}
%
\title{People Counting using Radio Irregularity in Wireless Sensor Networks -- An Experimental Study}
%
%
%
%

\author{
		Wei-Chuan~Lin, 
        Winston~K.G.~Seah 
        and~Wei~Li
\IEEEcompsocitemizethanks{
\IEEEcompsocthanksitem Wei-Chuan~Lin, and Winston~K.G.~Seah are with the School of Engineering and Computer Science, Victoria University of Wellington, P.O. Box 600, Wellington 6140, New Zealand. E-mail: \{wei-chuan.lin, winston.seah\}@ecs.vuw.ac.nz
\IEEEcompsocthanksitem Wei~Li is with the Department of Electrical and Computer Engineering, University of Victoria, P.O. Box 3055, Victoria, B.C. V8W~3P6, Canada. Email: wli@ECE.UVic.CA\protect\\
}
}

\IEEEcompsoctitleabstractindextext{%
\begin{abstract}
The Internet has grown into a large cyber-physical system centered that connects not just computer systems but a plethora of systems, devices, and objects, collectively referred to as ``Things'', giving rise to the term ``Internet of Things'' (IoT). It encompasses technologies for identification and tracking, sensing and actuation, both wired and wireless communications, and also, intelligence and cognition. Wireless communications, which is an integral part of IoT, suffers from radio irregularity -- a phenomenon referring to radio waves being selectively absorbed, reflected or scattered by objects in their paths, e.g., human bodies that comprises liquid, bone and flesh. Radio irregularity is often regarded as a problem in wireless communications but, with the envisioned pervasiveness of IoT, we aim to exploit radio irregularity as a means to detect and estimate the number of people. We demonstrate how radio signal fluctuations arising from radio irregularity, combined with discriminant analysis, can be used to provide a simple low-cost alternative to dedicated sensing systems for indoor people counting.
\end{abstract}

}

\maketitle

\IEEEdisplaynotcompsoctitleabstractindextext

%
\IEEEpeerreviewmaketitle

\section{Introduction}
%
%

%
%
%
%
\IEEEPARstart{T}{he} Internet has grown beyond connecting computer systems and platforms that run applications to meet endusers' computing and communication needs to connecting a plethora of systems, appliances, devices, objects, etc., collectively referred to as ``Things'', turning it into a large cyber-physical system and giving rise to a new paradigm known as the ``Internet of Things'' (IoT)~\cite{CERT-IoT:2009}. Likewise, the technologies that the IoT encompasses extend beyond computation and communication, to identifcation and tracking, sensing and actuation, and even intelligence and cognition. Ensuring connectivity in the IoT is increasingly reliant on wireless communications as connected devices become more ubiquitous and embedded into our daily living space.

The key technology in wireless communications is radio frequency (RF) transmission. When an RF signal propagates within a medium, it may be reflected, diffracted, and scattered. Each effect occurs to a different extent in various media, depending on factors such as wavelength and intensity of the wave, thickness and physical composition (permittivity and permeability) of the medium. The human body comprises liquid, bone and flesh, which selectively absorb, reflect or scatter RF signals, leading to the phenomenon known as \textsl{radio irregularity}. Consequently, in the presence of human activity within a network, the radio irregularity phenomenon is seen as signal strength fluctuations at the receiver, and the degree of signal fluctuation exhibits a significant level of correlation to the level of human activity in the network. Researchers have exploited this phenomenon in intrusion detection~\cite{Lee:MST2010} and device-free localization applications~\cite{Youssef:Mobicom2007,Patwari:PIEEE2010}.

Applications like automated people counting cannot tolerate false positives that result in overcounting, giving inaccurate data that are used for forecasting and resource allocation. People counting is extensively used in different industries, including retail (stores, malls and shopping centres), colleges and universities, government facilities, government non-profits organizations, visitor centres, libraries, museums and art galleries. In the retail industry, it is a form of intelligence-gathering that helps a retailer determine the percentage of visitors who actually make purchases. This is a key performance indicator of a store's performance as compared to just looking at the sales data. It also helps the management to optimize the usage of staff resources, e.g. deploy more staff during peak periods and cutting down during lull periods in order to save wages. For building automation and management purposes, people counting is used to optimize the use of resources as well as ensure that the safe level of occupancy is maintained.

With the emergence of IoT leading to pervasive wireless communication devices, radio irregularity which has often been viewed as a problem can instead be exploited for automated people counting with minimal additional hardware and installation costs. The goal of our approach is to provide a simple low-cost system for people counting that can be used independently or as a supplement to other more sophisticated methods. In the next section, we examine the related research on automated people counting with a focus on indoor use cases. We then present our approach to indoor automated people counting based on the signal fluctuations arising from radio irregularity. This is followed by the discussion of the experimental study and the application of discriminant analysis to process the results obtained from tests carried out indoors within a building before concluding the paper. 

%

\section{Related Work}
\subsection{People Counting Methods}
Widespread usage of people counting has fed the growth of commercially available automated people counting technology, among which infrared (IR) beam counters, thermal counters and video/CCTV cameras are most often used. 
The simplest and possibly cheapest approach is a single-beam IR counter placed across an entrance. However, such a counter suffers from numerous drawbacks and is only suitable detecting someone passing, e.g. entering/leaving a shop. These commonly used counters have very high percentage of errors when multiple persons cross their monitoring area at a time. When multiple (IR) beams or other forms of boundary sensors are deployed with careful placements strategies and coupled with artificial intelligence and/or analytical techniques for processing, a more accurate and versatile people counting system can be realized~\cite{Mathews:WISES2008,Zheng:TOSN2007}.

People counters that use thermal imaging are typically mounted overhead and have the ability to simultaneously maintain separate counts for multiple people moving in two directions (in and/or out). The IR images captured by the heat detectors are then processed to determine the number of people~\cite{Nakamura:IMTC1994}. Video-based people counters work on video streams obtained through video/CCTV camera which are then run through intelligent video-processing techniques to identify and count the people in the video~\cite{Teixeira:JSP2008}. The accuracy of such approaches can vary according to the level of ambient lighting and background colour contrasts~\cite{Celik:ICIP2006}. Hybrid approaches combining IR and video cameras, together with neural networks, have been proposed to improve the accuracy of visual-based automated people counting~\cite{Amin:Measurement2008}.

\subsection{Radio-based Detection and Counting Methods}
It was first reported in~\cite{Woyach:WiOpt2006} that the shadowing effect caused by an object moving between two communicating wireless devices can be used for detection purposes. In particular, a human body comprises liquid, bone and flesh, that selectively absorb, reflect or scatter RF signals, leading to the phenomenon known as \textsl{radio irregularity}. This phenomenon has been extensively used for device-free localization in wireless networks~\cite{Youssef:Mobicom2007,Patwari:PIEEE2010}. Radio Tomographic Imaging~\cite{10.1109/TMC.2009.174} measures the attenuation of signals across wireless links between many pairs of nodes in a wireless network to create ``images'' of objects moving within the network area. The variance of measured received signal strength (RSS) on the links in a network has also been used to infer the locations of people or objects moving in the network deployment area~\cite{Patwari:TIFS2011}. This approach utilizes a statistical model for the RSS variance as a function of a person's position with respect to the transmitter and receiver locations. The approach adopted by~\cite{Woyach:WiOpt2006} has also been extended in~\cite{Puccinelli:PerSens2011} for outdoor people counting by measuring the RSS level measured at the receiver. The reliance on (absolute) RSS values, however, has a drawback during deployment, which is the need to take into consideration the channel model and other related factors like the impact of path loss and fading. These approaches also require complex signal processing techniques and a calibration phase for each deployment environment, and this can significantly affect the ease of deployment. Our targetted application differs from the proposed device-free localization schemes~\cite{Youssef:Mobicom2007,Patwari:PIEEE2010} as we are not aiming to localize or track individual objects.

It has been observed in~\cite{Lee:MST2010} that human movement through the path of the radio signal causes the histogram of the absolute RSS values to become more spread; this is manifested quantitatively as higher standard deviation. However, the standard deviation varies significantly across environments, making it difficult to define a universal threshold to detect movement in terms of these first order statistics. While also exploiting the RSS spread caused by human movement, the approach adopted focused on the fluctuation in signal strength instead, in order to reduce the impact of channel models and other environmental factors. However, there are false positives reported in their results which are deemed to be acceptable for intrusion detection applications. This has since been improved by eliminating the false positives and extended to detect two persons crossing a narrow corridor~\cite{Lin:PIMRC2011}.

\section{Counting People using RSSI Fluctuations}
\subsection{RSSI fluctuations caused by human activity}

\begin{figure}[hbt]
\centering
\includegraphics[width=0.9\columnwidth,trim=0mm 7mm 0mm 4mm]{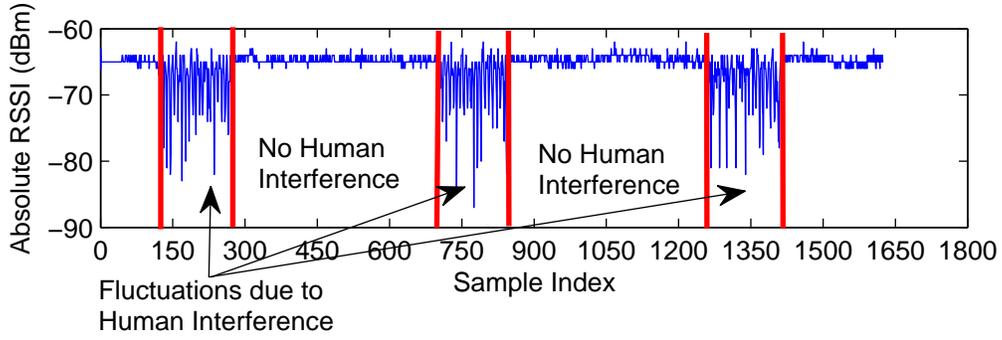}
\caption{Absolute RSSI reading}
\label{fig_Absolute_RSSI}
\end{figure}

\begin{figure}[hbt]
\centering
\subfloat[RSSI Fluctuation without movement]{
\includegraphics[width=0.9\columnwidth,trim=0mm 56mm 0mm 8mm,clip]{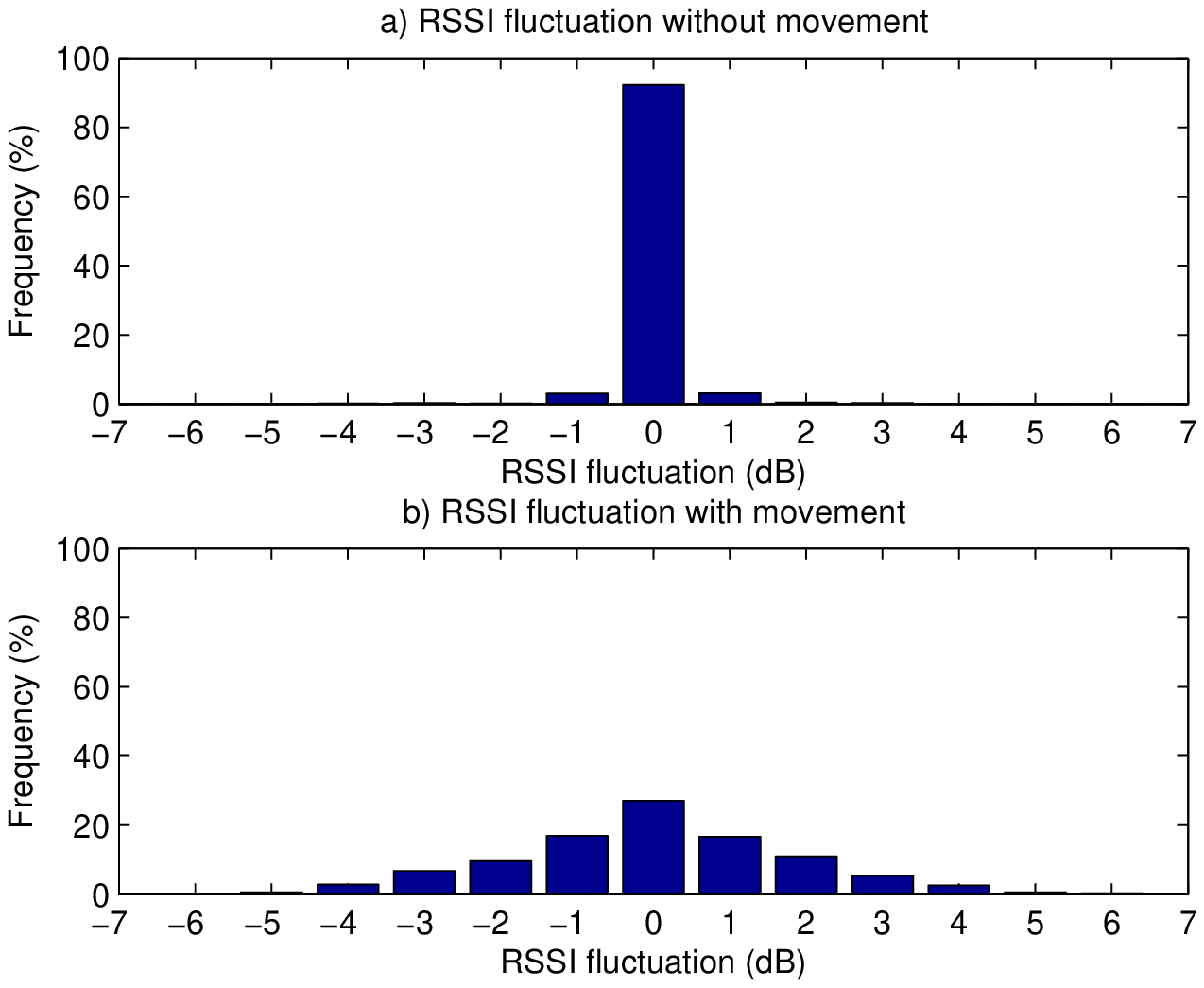}
\label{fig_RSSI_Fluctuation_Histogram_wo_movement}}
\\
\subfloat[RSSI Fluctuation with movement]{
\includegraphics[width=0.9\columnwidth,trim=0mm 3mm 0mm 60mm,clip]{RSSI_Fluctuation_Histogram}
\label{fig_RSSI_Fluctuation_Histogram_with_movement}}
\caption{RSSI Fluctuation Patterns~\cite{Lee:MST2010}}
\label{fig_RSSI_Fluctuation_Histogram}
\end{figure}

Most, if not all, the approaches that use the variance of RSS caused by human motion across the signal transmission paths require complex signal processing techniques and a calibration phase for each deployment environment, and this can significantly affect the ease of deployment. 
In our scheme, we adopt a network-oriented approach that relies on RSS information of received packets which can be easily obtained from device drivers of wireless network interfaces when the packets are received. A key goal of our approach is to be able to easily utilize the existing wireless transmitters and receivers already deployed in the environment without the need for accurate channel models nor complex signal processing techniques. We extend the method of using Received Signal Strength Indicator (RSSI) fluctuations proposed in~\cite{Lee:MST2010} where the RSSI fluctuation for a packet $p_i$, denoted by $F(p_i)$, is given by $F(p_i) = RSSI(p_i) - RSSI(p_{i-1})$. It has been shown that two consistent patterns of RSSI fluctuations can be observed for two key scenarios of interest to us, namely, without human movement and with human movement across the signal transmission path.  
Given the absolute RSSI readings for packets recorded at the receiver over 
time (Fig.~\ref{fig_Absolute_RSSI}), the histogram of RSSI fluctuations shows narrower distribution when there is no human movement across the signal path, i.e., less fluctuation across RSSI readings (Fig.~\ref{fig_RSSI_Fluctuation_Histogram_wo_movement}) and, conversely, signals fluctuate more in the presence of human movement resulting in the spread out distribution shown in  Fig.~\ref{fig_RSSI_Fluctuation_Histogram_with_movement}. 

\subsection{Human Detection}
\begin{figure}[hbt]
\centering
\includegraphics[width=0.9\columnwidth,trim=0mm 5mm 0mm 3mm]{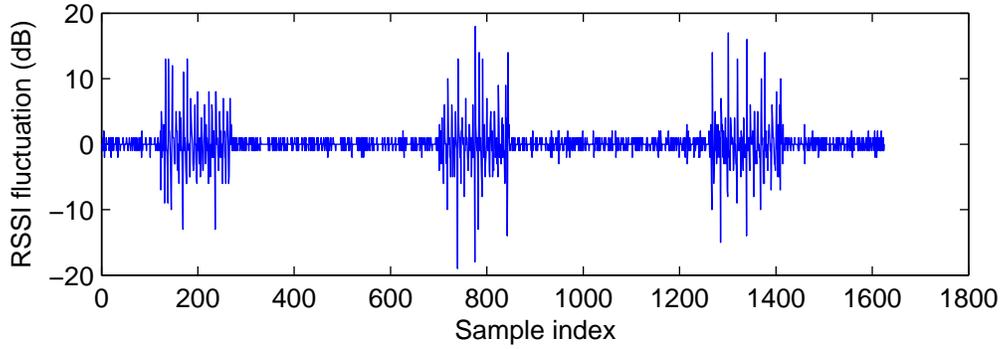}
\caption{RSSI Fluctuation}
\label{fig_RSSI_Fluctuation}
\end{figure}

Our proposed algorithms computes the fluctuation between the RSSI of packets received at a receiver since the absolute RSSI largely depends on external environment. From the absolute RSSI readings for packets recorded at the receiver over a period of time as shown in Fig.~\ref{fig_Absolute_RSSI}, the fluctuation of RSSI readings is calculated, as shown in Fig.~\ref{fig_RSSI_Fluctuation}. 

\begin{figure}[hbt]
\centering
\includegraphics[width=0.75\columnwidth, trim=2mm 5mm 2mm 5mm]{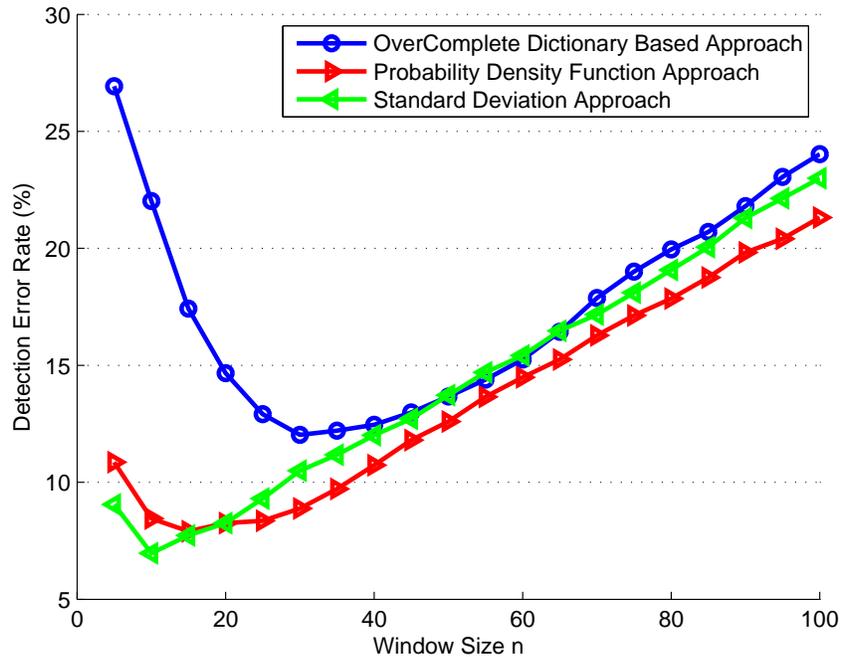}
\caption{Optimal Window Size}
\label{fig_Optimal_WindowSize_BER}
\end{figure}

\begin{figure}[hbt]
\centering
\includegraphics[width=0.8\columnwidth, trim=0mm 5mm 0mm 3mm]{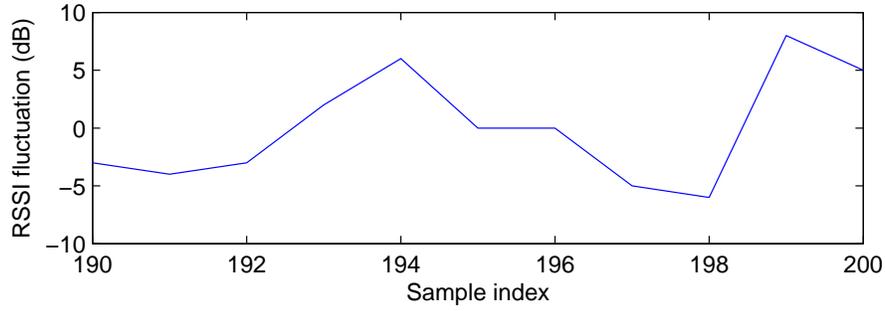}
\caption{RSSI fluctuations over a window size of 10}
\label{fig_Sample_Window}
\end{figure}

We define a sliding window of $n$ samples, where $n$ is a parameter that can be tuned to achieve the desired accuracy for the target environment. The desired value for $n$ would be one that achieved the lowest error detection rate, which is calculated by dividing the number of incorrect detections by the total number of samples. We studied the accuracy of detection using different sliding window sizes, from $n= 5$ to $n=100$, across different environments, and also comparing with other schemes, namely, the probability density function approach~\cite{Lin:PIMRC2011} and the overcomplete dictionary approach~\cite{Lin:PACRIM2011}, to obtain the results shown in Fig.~\ref{fig_Optimal_WindowSize_BER}. 
Based on the results shown in Fig.~\ref{fig_Optimal_WindowSize_BER}, we use a sliding window of size $n = 10$ to observe the behaviour of RSSI fluctuation. Therefore, a window of RSSI fluctuations at sample 200 is shown in Fig.~\ref{fig_Sample_Window}. 

\begin{figure}[H]
\centering{
\subfloat[Sample 200 (movement)]{\includegraphics[width=0.5\columnwidth, trim=3mm 0mm 2mm 0mm]{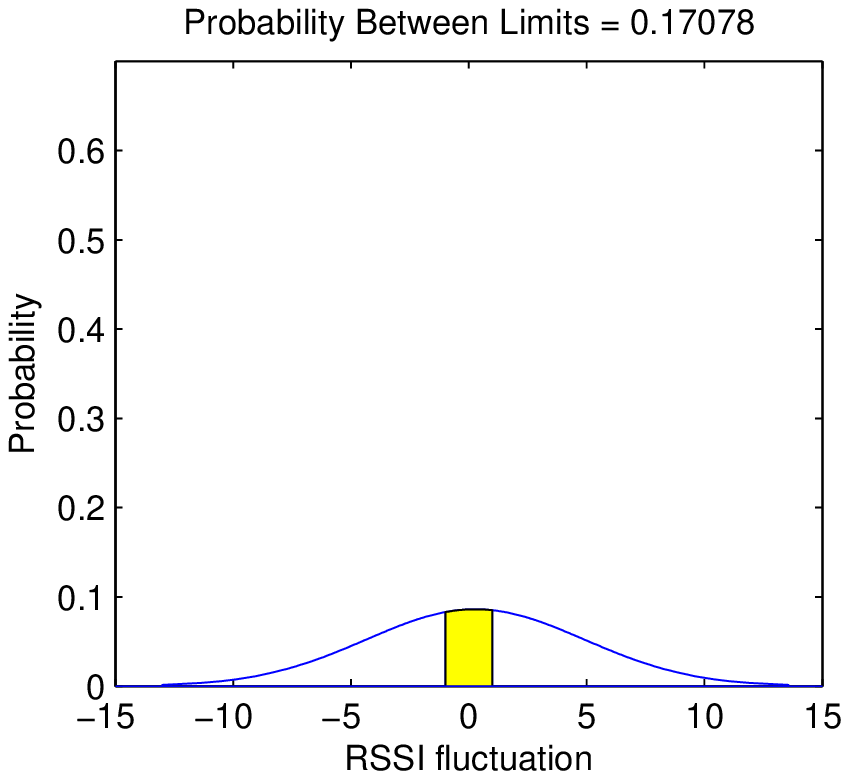}%
\label{fig_Normal_With_Movement_at_200}}
\hfil
\subfloat[Sample 600 (no movement)]{\includegraphics[width=0.49\columnwidth, trim=3mm 0mm 3mm 0mm]{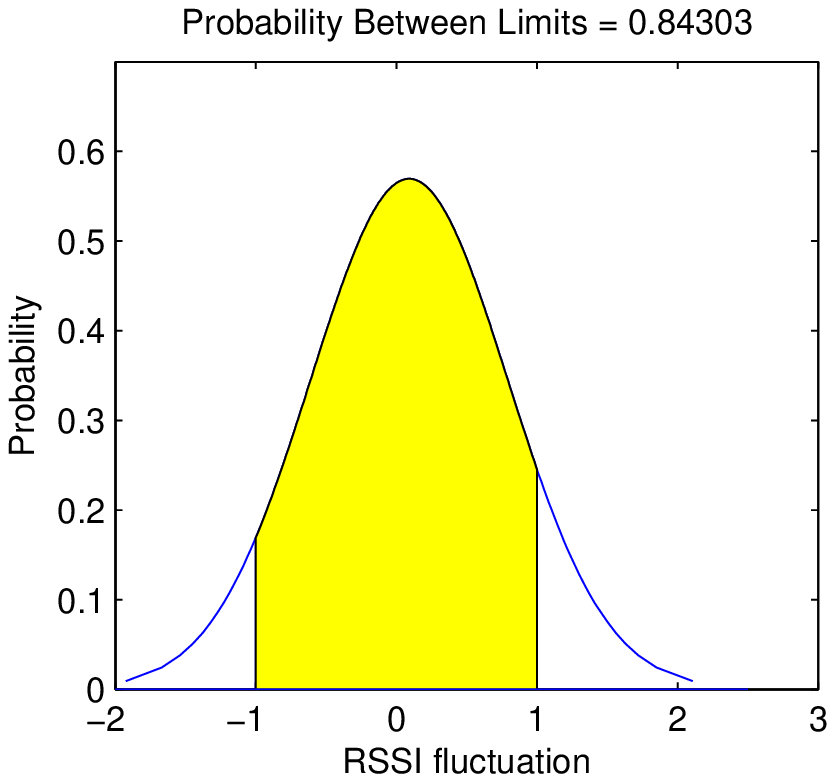}%
\label{fig_Normal_Without_Movement_at_600}}
}
\caption{Normal distribution showing probability in fluctuation range [-1,1]}
\label{fig_Normal_Distribution}
\end{figure}

At sample 200, using the window of 10 previous readings, the mean and standard deviation are computed as 0.2727 and 4.6280 respectively. We then map the RSSI fluctuations into the normal distribution with the mean and standard deviation for that window, i.e. $\mu = 0.2727$ and  $\sigma = 4.6280$, as shown in Fig.~\ref{fig_Normal_With_Movement_at_200} representing the case where the signal has been subjected to interference by human movement across its path. Similarly, the normal distribution of RSSI fluctuation at sample 600, where there is no movement, is shown for comparison in Fig.~\ref{fig_Normal_Without_Movement_at_600}. 

\begin{figure}[hbt]
\centering
\includegraphics[width=0.8\columnwidth]{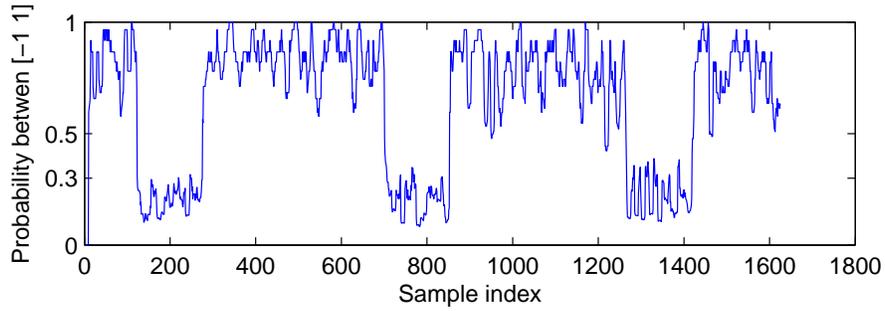}
\caption{Probability of fluctuation within [-1,1] for RSSI readings in Fig.\ref{fig_Absolute_RSSI}}
\label{fig_Probability_plus_minus_one_for_dataset}
\end{figure}

\begin{figure}[H]
\centering
\includegraphics[width=0.8\columnwidth,trim=0mm 8mm 0mm 2mm]{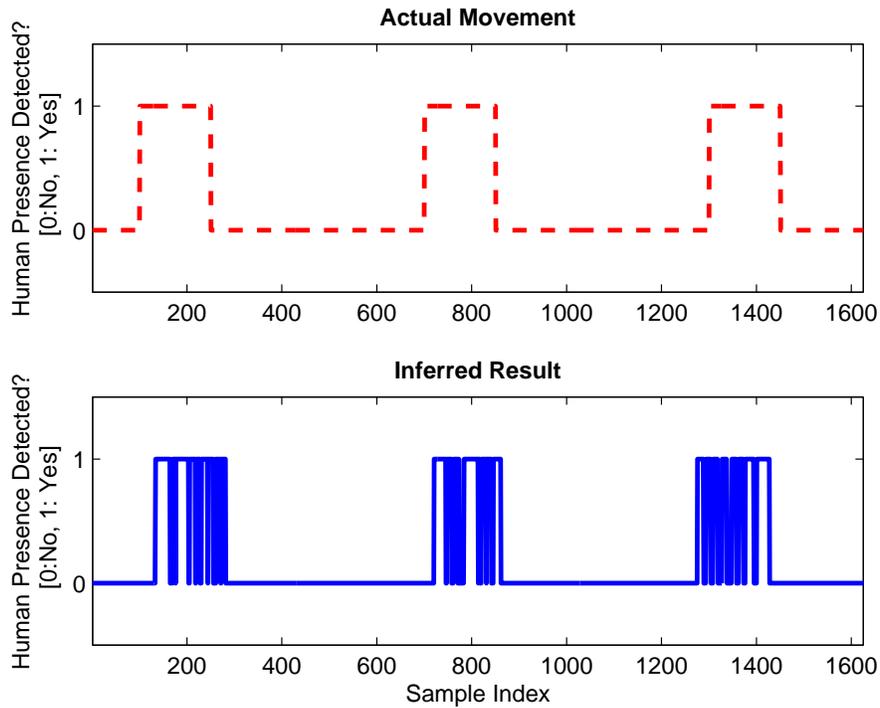}
\caption{Inferred presence of human movement using RSSI Fluctuations}
\label{fig_Detection_Output}
\end{figure}

From the graphs, we compute the probability of the RSSI fluctuation falling within the range [-1,1] (i.e. area under the curve from -1 to 1) to be 0.17078 for the case where there is movement across the signal path (i.e. sample 200) and 0.84303 for the case where there is no movement (sample 600). For the dataset shown in Fig.~\ref{fig_Absolute_RSSI}, we compute the probability of falling with the fluctuate range [-1,1] and plot the results 
in Fig.~\ref{fig_Probability_plus_minus_one_for_dataset}.
As shown, the probability of fluctuations falling in the range of [-1, 1] is below 0.3 in the presence of human movement. Hence, a probability value that is higher than 0.3 implies no human movement. Based on this threshold, we then infer from the results whether or not there has been human movement across the signal path, and the results are shown in Fig.~\ref{fig_Detection_Output}. 

\begin{figure}[hbt]
\centering{
\subfloat[False positives by detection method in \cite{Lee:MST2010}]{\includegraphics[width=0.9\columnwidth,trim=0mm 2mm 0mm 10mm]{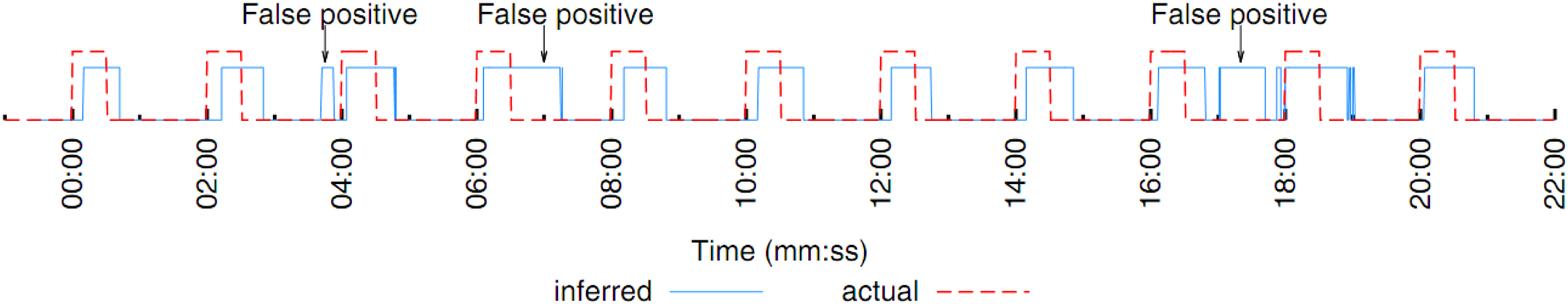}
\label{fig_False_Positive}}
\hfil
\subfloat[Movement detected by our method]{\includegraphics[width=0.9\columnwidth,trim=20mm 2mm 15mm 8mm,clip]{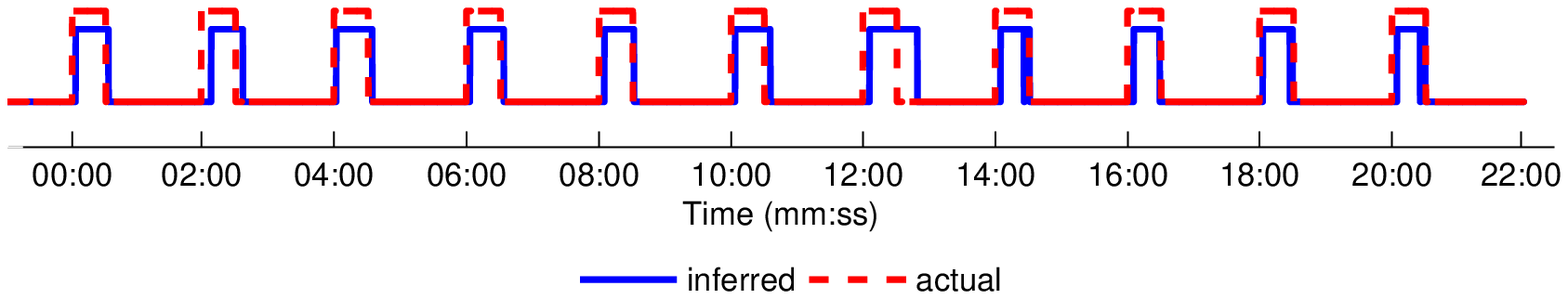}
\label{fig_Detection_Using_PIMRC2009_dataset}}
\caption{Detection results using dataset of \cite{Lee:MST2010}}
\label{fig_Our_Results}}
\end{figure}

The approach used in~\cite{Lee:MST2010} has resulted in false positives as shown in Fig.~\ref{fig_False_Positive}.
We applied our approach to the dataset used by the detection algorithm~\cite{Lee:MST2010} that produced the results shown in Fig.~\ref{fig_False_Positive}, and confirmed that our algorithm is able to achieve better accuracy in eliminating false positives, as shown in Fig.~\ref{fig_Detection_Using_PIMRC2009_dataset}.

\section{Pedestrian Traffic Monitoring}
Next, we extend the detection algorithm to monitor pedestrian traffic with the aim of counting the number of people passing through a particular location of interest.
\subsection{Single transmitter-single receiver configuration}
\begin{figure}[hbt]
\centering
\includegraphics[width=0.7\columnwidth]{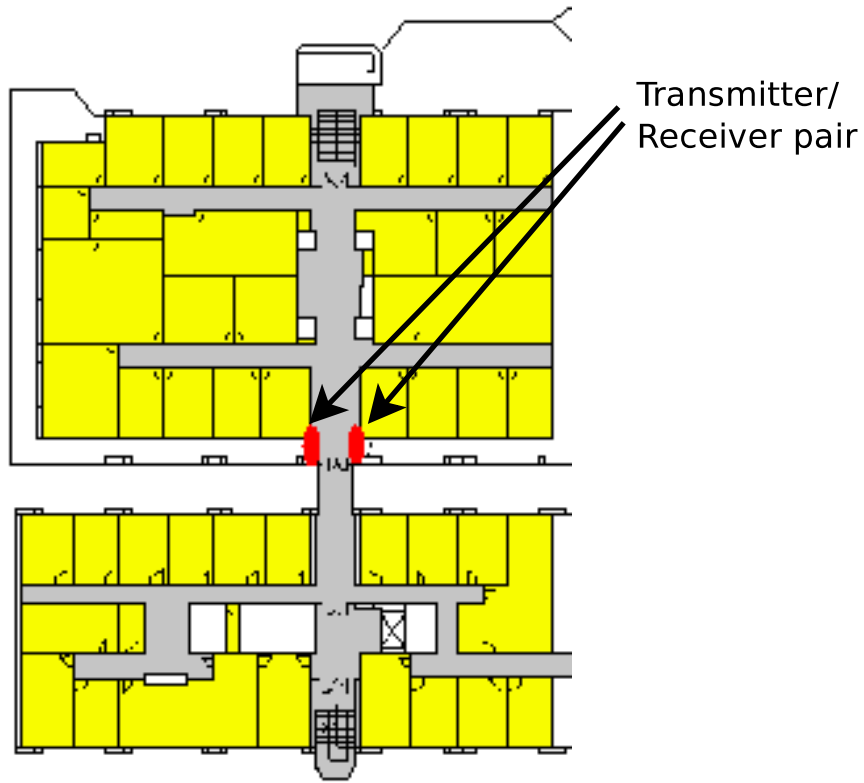}
\caption{Deployment along corridor of building in university}
\label{fig_Deployment_Site}
\end{figure}

\begin{figure}[hbt]
\centering
\includegraphics[width=0.9\columnwidth,trim=0mm 0mm 0mm 0mm,clip]
{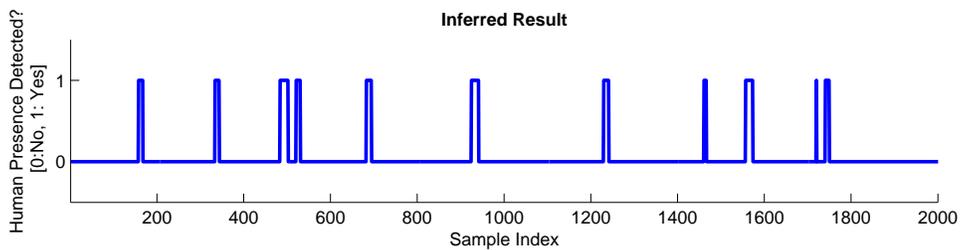}
\caption{Detection of pedestrian traffic along corridor}
\label{fig_Detection_At_Corridor}
\end{figure}

First, a series of experiments were conducted to observe the precision of the detection algorithm in a realistic indoor environment, namely, a corridor in a university building, as shown in Fig.~\ref{fig_Deployment_Site}, where the two red dots indicated by the arrows refer to the transmitter/receiver pair using IEEE802.15.4 technology. The devices are spaced 1.5m apart (width of corridor) and placed at a height of 1.1m, on a ledge. Each data collection duration was 300 seconds with inter-packet interval time of 0.15 seconds, during which the number of people who have walked past the devices were recorded and tagged with the time. Fig.~\ref{fig_Detection_At_Corridor} shows the results for one data collection period, during which nine persons walked through individually and two pairs of people past while walking close to each other, at the sample index of 484 and 925. In the detection results, shown in Fig.~\ref{fig_Detection_At_Corridor}, a total of 11 movements were detected. It is clear that detecting two people walking side by side is a major challenge as the RSSI fluctuations arising from one and two persons passing are quite indistinguishable.

\subsection{Single transmitter-multiple receiver configuration}
\begin{figure}[hbt]
\centering{
\subfloat[Conceptual Configuration]{\includegraphics[width=0.45\columnwidth]{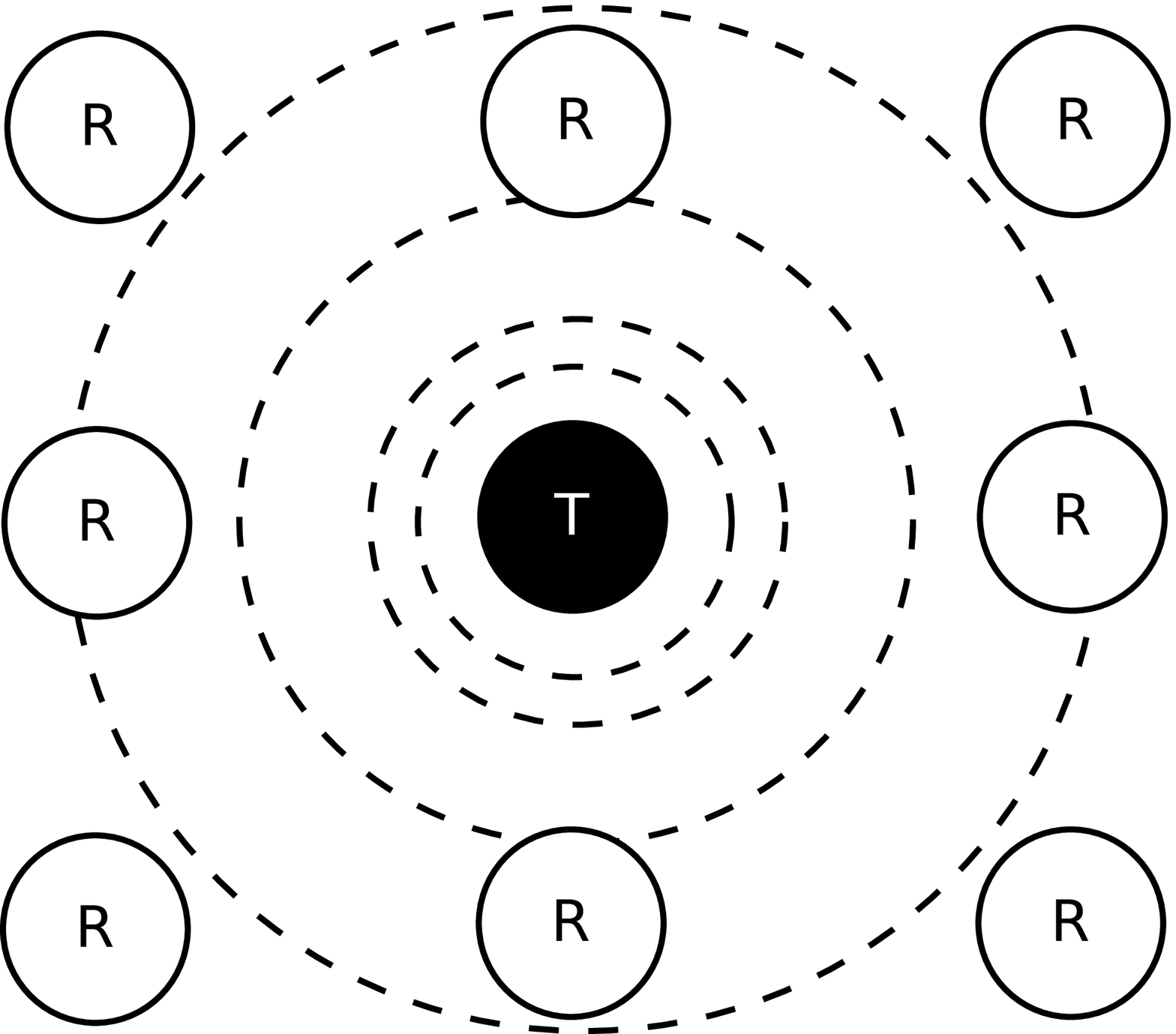}
\label{fig_One-Txr-Many-Rxr}}
\hfil
\subfloat[One-Transmitter Two-Receiver]{\includegraphics[width=0.52\columnwidth]{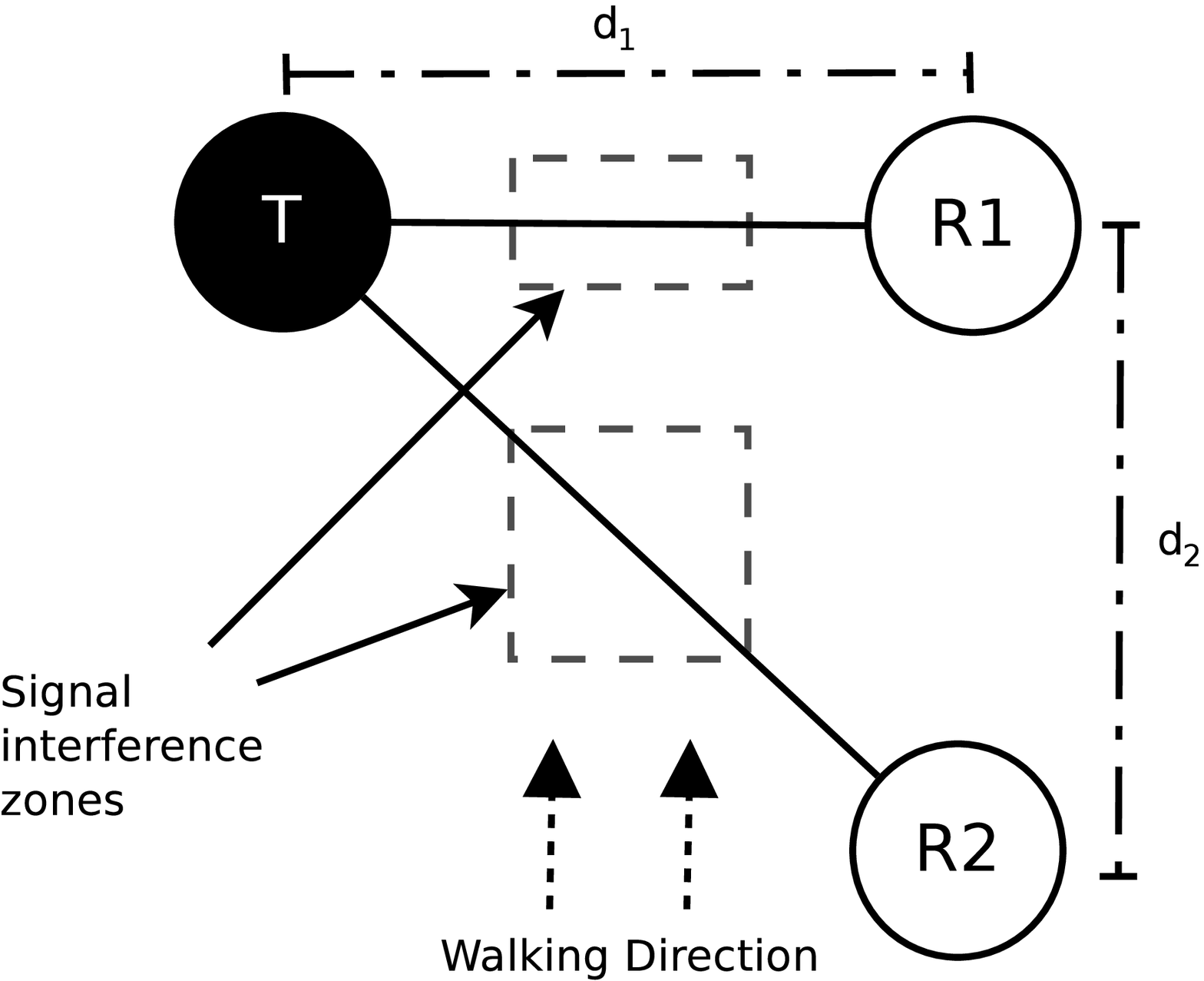}
\label{fig_One-Txr-Two-Rxr}}
\caption{Single Transmitter with Multiple Receivers}
\label{fig_Single_Transmitter_with_Multiple_Receivers}}
\end{figure}

A cyber-physical system like IoT is made up of numerous small wireless devices, from which a conceptual deployment scenario like that shown in Fig.~\ref{fig_One-Txr-Many-Rxr} can be assumed; we first look at a subset configuration of one-transmitter and two-receivers as shown in Fig.~\ref{fig_One-Txr-Two-Rxr}~\cite{Lin:PIMRC2011}. As people walk along the path between T and the two receivers in the direction shown in Fig.~\ref{fig_One-Txr-Two-Rxr}, they first cross the T-R$_2$ signal transmission path, followed by the T-R$_1$ signal path. A key point to note is the different signal interference zones that result from the movement of the people. 

Using the one-transmitter two-receiver configuration, the transmitter broadcasts packets at a rate of one packet every 0.15 seconds. Receiver R$_1$ is 1.5m from transmitter T (d$_1$ = 1.5m) and R$_2$ is 1.5m from  R$_1$ (d$_2$ = 1.5m). As two persons walk along the path between T and the two receivers in the direction shown in Fig.~\ref{fig_One-Txr-Two-Rxr}, they first cross the T-R$_2$ signal transmission path, followed by the T-R$_1$ signal path.

\begin{figure}[hbt]
\centering
\includegraphics[width=0.8\columnwidth,trim=0mm 2mm 0mm 2mm]{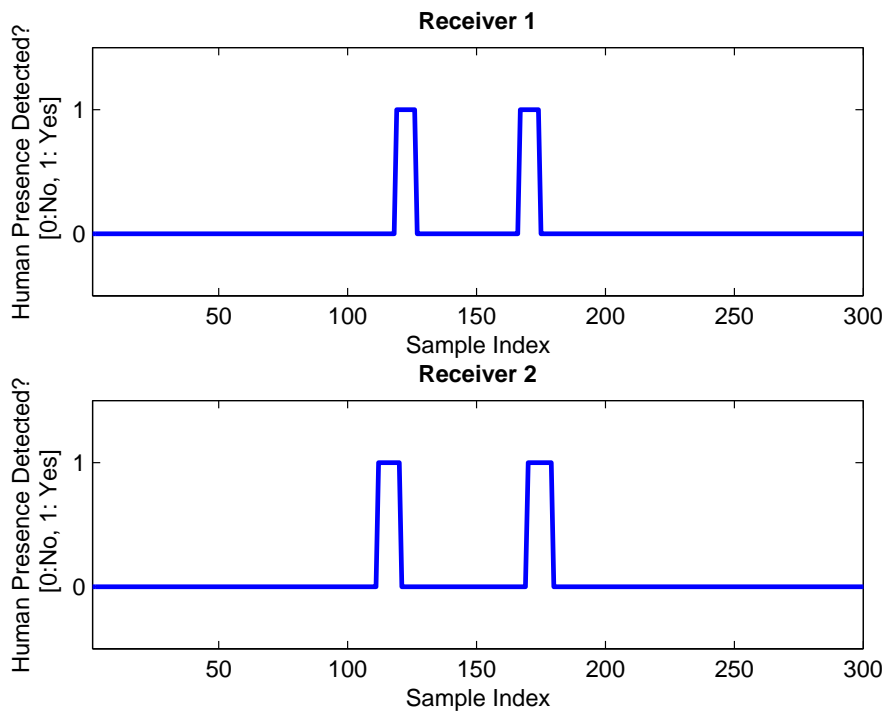}
\caption{One person walking in the direction of R2 to R1}
\label{fig_One-Person-R2-to-R1}
\end{figure}

First, we collected data for one person walking across the signal transmission path, passing first R$_2$ then R$_1$ to be used as the reference case. The detection results correctly show that one person passed at around the time of sample 100 and another at around sample 200, as shown in Fig.~\ref{fig_One-Person-R2-to-R1}. Intuitively, the detection result at sample 100 is more logical since the person passed R$_2$ first, then R$_1$. However, as the two receivers at very close to each other, having the two receivers showing signal fluctuations at almost the same instant is also likely especially when the person is walking fast.

\begin{figure}[hbt]
\centering
\includegraphics[width=0.8\columnwidth,trim=0mm 2mm 0mm 2mm]{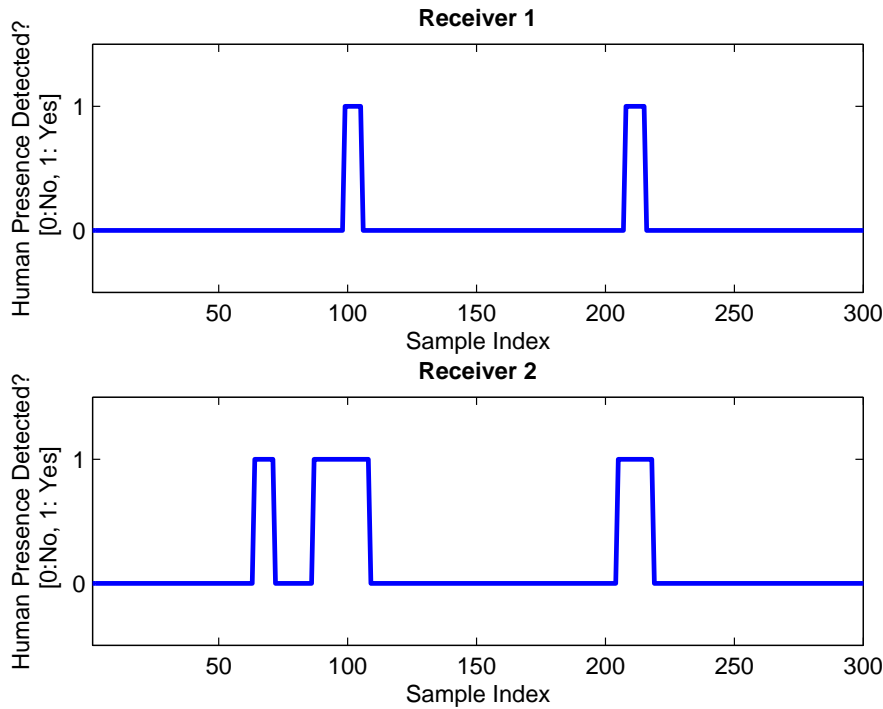}
\caption{Two people walking in the direction of R2 to R1}
\label{fig_Two-Persons-R2-to-R1}
\end{figure}

\begin{figure}[hbt]
\centering
\includegraphics[width=0.8\columnwidth,trim=0mm 2mm 0mm 2mm]{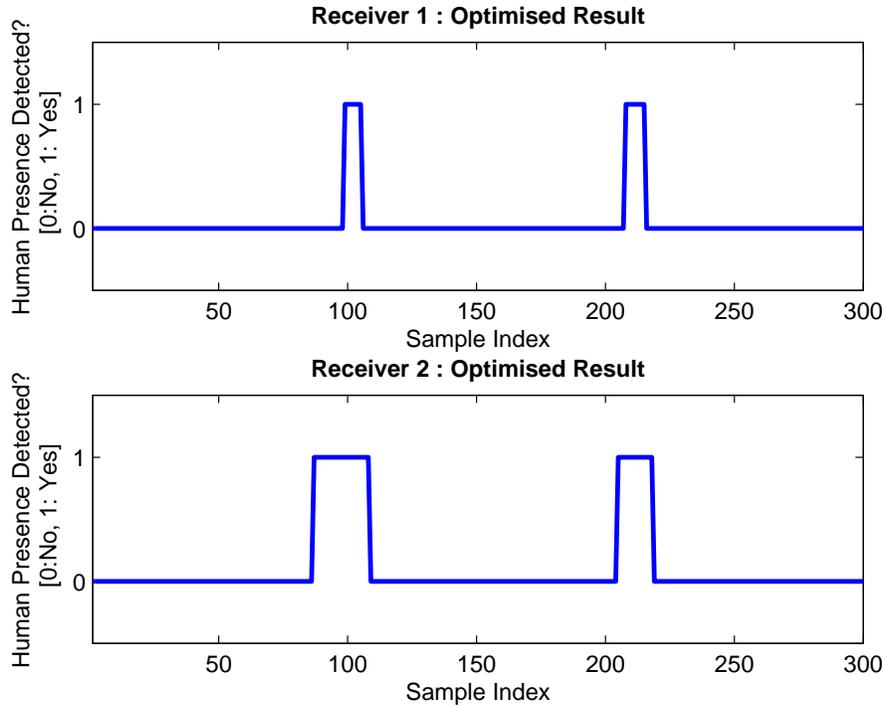}
\caption{Optimised Result of multiple receivers}
\label{fig_Two-Persons-R2-to-R1-optimized}
\end{figure}

Next, we collected data for the case of two persons walking side-by-side in the direction of R$_2$ to R$_1$ as shown in Fig.~\ref{fig_One-Txr-Two-Rxr}. We expect that the detection duration of T-R$_2$ should be longer than T-R$_1$. This is because the T-R$_2$ signal experienced a longer duration of interference than the T-R$_1$ signal. The detection result of two people walking from R$_2$ to R$_1$  shown in Fig.~\ref{fig_Two-Persons-R2-to-R1} confirms our hypothesis.
However, we also observed a false positive detection at sample 64. As the two receivers are placed closed to each other, 1.5m apart, we can assume that it is unlikely for a moving object to be detected by one receiver but not the other. Therefore, by comparing and matching the data from both receivers, we can perform a simple optimization process to remove such false positive detections, to achieve the desired results as shown in Fig.~\ref{fig_Two-Persons-R2-to-R1-optimized}.


\section{Enhanced Accuracy for People Counting}
\subsection{Standard Deviation of RSSI fluctuation Approach}
Using the probability density function of RSSI fluctuations falling within the range [-1,+1] as the threshold to signify no human movement has eliminated the occurrence of false positives~\cite{Lin:PIMRC2011}, but it disregards information from the distribution of RSSI fluctuations that lie outside the region between -1 and +1. This is undesirable as the distribution of RSSI fluctuations has been shown to be a good indication of the size or crowd density of moving objects~\cite{Nakatsuka2008}.

\begin{figure}[hbt]
\centering
\includegraphics[width=0.8\columnwidth, trim=0mm 2mm 0mm 3mm]{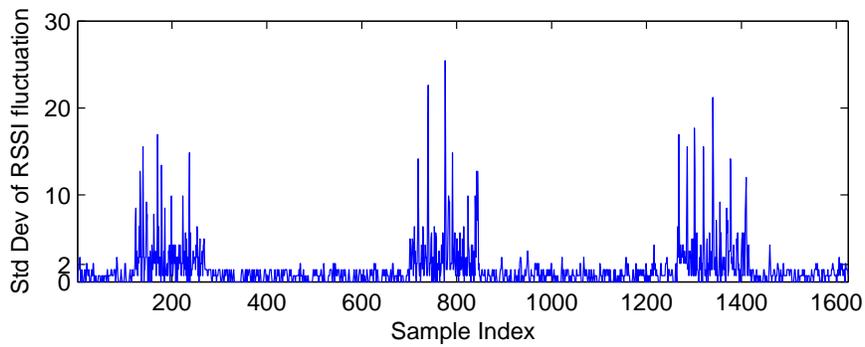}
\caption{Standard Deviation of fluctuation for RSSI readings in Fig.\ref{fig_Absolute_RSSI}}
\label{fig_Standard_deviation_for_dataset}
\end{figure}

\begin{figure}[hbt]
\centering
\includegraphics[width=0.8\columnwidth, trim=0mm 5mm 0mm 3mm]{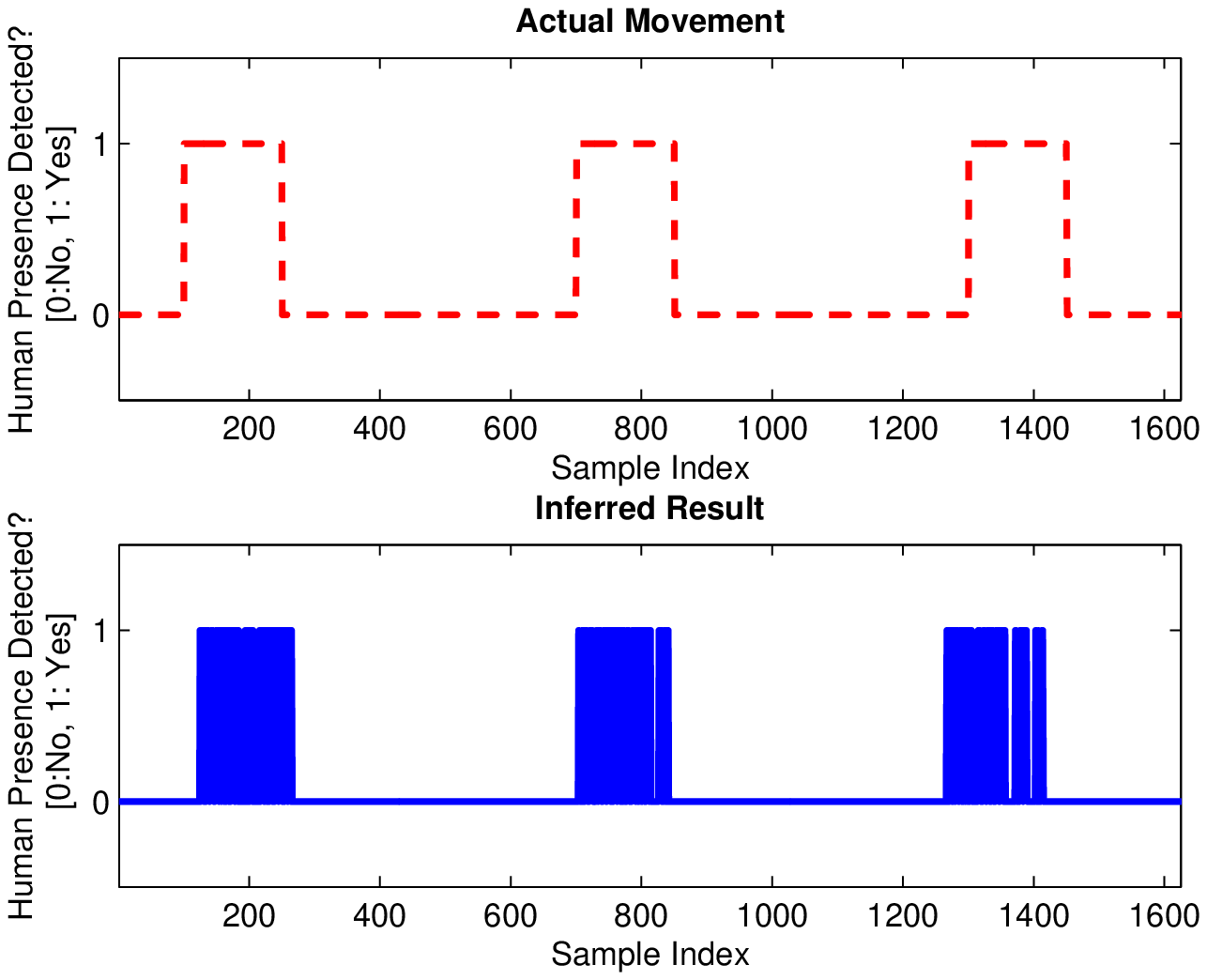}
\caption{Inferred presence of human movement using standard deviation}
\label{fig_Detection_Output_Std}
\end{figure}

According to Fig.~\ref{fig_Optimal_WindowSize_BER}, we use a sliding window of size $n = 10$ to observe the behaviour of RSSI fluctuation. Signal interference due to human motion causes rapid RSSI fluctuations which results in an increased standard deviation. 
For the dataset shown in Fig.~\ref{fig_Absolute_RSSI}, we compute the standard deviation of RSSI fluctuation and the computed results are shown in Fig.~\ref{fig_Standard_deviation_for_dataset}; at sample 200 (human presence), the standard deviation (of the most recent 10 RSSI fluctuation readings) is 4.6280 and at sample 600 (no human presence), the standard deviation is 0.6325. 
The standard deviation of RSSI fluctuations being higher than 2 implies the presence of human movement, which provides a clearer threshold than the approach presented in~\cite{Lin:PIMRC2011}. We infer the existence of human movement based on the standard deviation threshold and the results are shown in Fig.~\ref{fig_Detection_Output_Std}. More importantly, additional information can be derived from the data, such as, the peak of standard deviation.
 
\subsection{Multiple Object Detection Setup}
In order to exploit the effect of RSSI fluctuation caused by signal interference by a few people, we conducted a series of experiments with a variable group size (ranging from one to five people) under a controlled indoor environment.  Sensor motes using IEEE802.15.4 wireless technology were deployed in a $6m{\times}8m$ room in a one-transmitter two-receiver configuration (Fig.~\ref{fig_One-Txr-Two-Rxr}.) The motes were placed at a height of 0.9m. Receiver R$_1$ was placed 3.5m from transmitter T (d$_1$ = 3.5m) and R$_2$ at 2m from R$_1$(d$_2$ = 2m). The transmitter, T, broadcasts packets continuously in time intervals of 0.15s. The absolute RSSI values were recorded upon packet reception at the receivers. Then, groups comprising one to five people made five consecutive round trips between the transmitter and two receivers. Five experiments were conducted for each group. 

\subsection{Data Analysis}
\begin{figure}[hbt]
\centering
\includegraphics[width=0.8\columnwidth,trim=0mm 7mm 0mm 0mm]{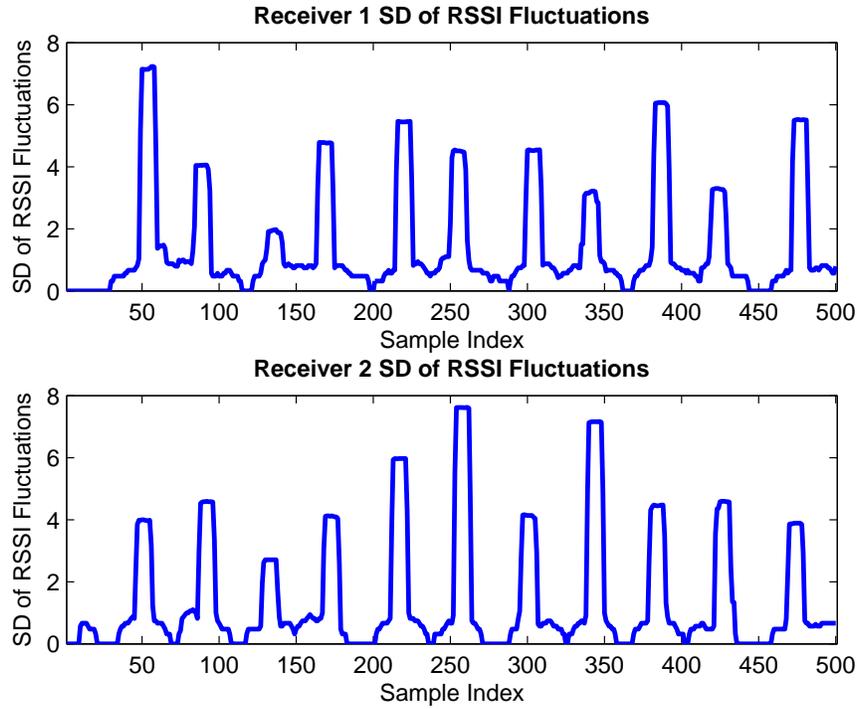}
\caption{Experimental Data for one person in a single experiment}
\label{fig_Experimental_Data}
\end{figure}

We use \textit{discriminant analysis}, a method to find the linear combination of measurements which characterize two or more groups \cite{McLachlan1993}, to analyze the collected data. 
The key concept of the \textit{discriminant analysis} approach is to classify the number of people based on the difference in the influence on RSSI readings due to the size of interference zone.
We let a finite number $g$ denote the distinct number of groups which in our case is five (i.e. $g$=5, denoting groups of 1--5 persons). We refer to the $G_i$ as groups where $i = 1, \ldots, 5$. There are two phases of discriminant analysis which are \textit{training} and \textit{classification}. The system identifies  differences in RSSI fluctuations caused by different number of people as signatures in the training phase. It is important to note that this training is only required once to identify the fluctuations caused by the different number of people, and not for each deployment scenario. During the training phase, a total of $g-1$ orthogonal discriminant functions are constructed such that the groups differ as much as possible on discriminant score $D$. The form of the linear discriminant function is: 
\begin{equation}
D = v_1 X_1 + v_2 X_2+ ... +v_i X_i+c
\end{equation}
where
\begin{align*}
 v &= \mathrm{the\ discriminant\ coefficient}\\
 X &= \mathrm{the\ value\ of\ independent\ variable}\\
 c &= \mathrm{a\ constant}\\
 i &= \mathrm{the\ number\ of\ independent\ variables}
\end{align*}
Once the discriminant functions are constructed, the discriminant analysis enters the second phase which is classification. Fisher's linear discriminant analysis~\cite{AHG:AHG2137} is used for data classification; the purpose of Fisher's technique is to find the line of projection that separates different groups~\cite{Welling2005}.
We use the standard deviation of RSSI fluctuations of detected movement as the primary dataset. For instance, the standard deviation of RSSI fluctuations of an experiment involving one person is shown in Fig.~\ref{fig_Experimental_Data}. We utilize the information from each positive detection, particularly mean, standard deviation (std), coefficient of variation (CV), detection duration and area under the curve, to be the independent variables in order to achieve high discrimination between groups. The more statistical information we can extract from these positive detections, the greater the ability to discriminate between the different size groups.

\begin{figure}[H]
\centering
\includegraphics[width=0.9\columnwidth,trim=0mm 5mm 0mm 0mm]{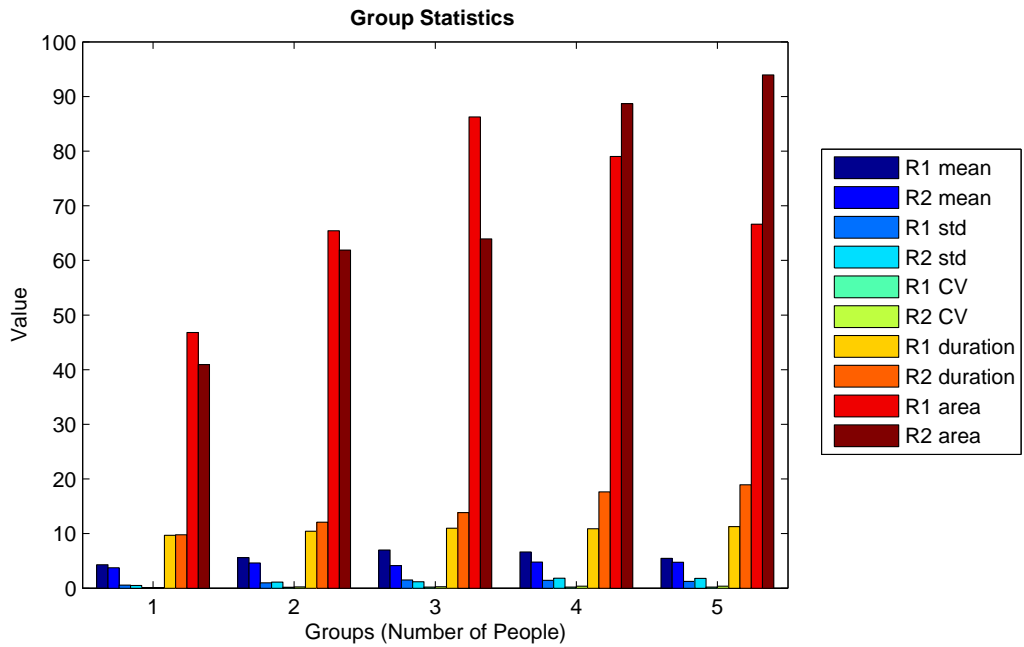}
\caption{Mean of independent variables of each group}
\label{fig_Group_Statistics}
\end{figure}

For a total of 50 samples of each group, the mean of each independent variable is plotted in Fig.~\ref{fig_Group_Statistics}. Duration and area under the curve are the two most significant independent variables. The detection duration of R$_1$ is stable throughout all groups while the detection duration of R$_2$ steadily increases as the number of people increase. This trend is as expected for more time is 
\noindent needed to pass the T-R$_2$ transmission path than T-R$_1$. The area under the curve follows the same trends as longer detection duration makes the area under the curve larger. 

Next, these independent variables are taken to construct the discriminant functions that maximize the separation between each group. In Table~\ref{tab_Test_of_Equality}, it provides very high F values and lower Wilks' Lambda as evidence of significant difference in R$_2$ detection duration than any other independent variables. In addition, in statistical significance testing, the null hypothesis is rejected when the $p$-value is smaller than the significance level $\alpha$ which is 0.05. The results are considered to be statistically significant when null hypothesis is rejected.
The information of each discriminant function is shown in Table~\ref{tab_Table_of_Eigenvalues}. There are five groups, namely `one person' to `five people' and as a result four discriminant functions are produced. With high Eigenvalue and percentage of variance, `Function 1' covers most total statistical population. 
\begin{table}[hbt]
\begin{center}
\begin{tabular}{lcccc}\hline
Independent Variable & Wilks' Lambda & F & $p$-value\\
 \hline
\hspace{2em}R$_1$ mean & 0.773 & 17.986 & 0.001\\
\hspace{2em}R$_2$ mean & 0.992 & 5.207 & 0.000\\
\hspace{2em}R$_1$ std & 0.640 & 34.437 & 0.000\\
\hspace{2em}R$_2$ std & 0.692 & 27.311 & 0.000\\
\hspace{2em}R$_1$ CV & 0.485 & 65.034 & 0.000\\
\hspace{2em}R$_2$ CV & 0.436 & 79.137 & 0.000\\
\hspace{2em}R$_1$ duration & 0.633 & 35.572 & 0.000\\
\hspace{2em}R$_2$ duration & 0.256 & 178.008 & 0.000\\
\hspace{2em}R$_1$ area & 0.713 & 24.605 & 0.000\\
\hspace{2em}R$_2$ area & 0.603 & 40.256 & 0.000\\\hline
\end{tabular}
\caption{Test of Equaltiy of Group Means}
\label{tab_Test_of_Equality}
\end{center}
\end{table}
\vspace*{-8ex}
\begin{table}[hbt]
\begin{center}
\begin{tabular}{cccc}\hline
Functions & Eigenvalue & \% of variance & Canonical Correlation\\
 \hline
1 & 5.717 & 82.9 & 0.923\\
2 & 0.952 & 13.8 & 0.698\\
3 & 0.146 & 2.1 & 0.357\\
4 & 0.084 & 1.2 & 0.279\\ \hline
\end{tabular}
\caption{Table of Eigenvalues}
\label{tab_Table_of_Eigenvalues}
\end{center}
\end{table}

The relative importance of each independent variable in each discriminant function can be found by analyzing the structure matrix table (Table~\ref{tab_Structure_Matrix}). These `Pearson' coefficients are discriminant loadings which act like factor loadings in factor analysis~\cite{Burns2008}. Generally, an absolute value of 0.3 is used as the threshold to separate a significant from an insignificant variable. For example, we have five independent variables in discriminant function $D_1$, namely, `R$_1$ CV', `R$_2$ CV', `R$_1$ duration', `R$_2$ duration', and `R$_2$ area' that discriminates between groups. The discriminant function coefficient shows the contribution of each independent variable to the discriminant function (Table~\ref{tab_Discriminant_Function_coefficient}). It operates like a regression equation. 

\begin{table}[hbt]
\begin{minipage}[b]{.45\textwidth}
\begin{center}
\begin{tabular}{lcccc} \hline
& \multicolumn{4}{c}{Function} \\ \cline{2-5}
& 1 & 2 & 3 & 4 \\ \cline{1-5}
R$_1$ mean & 0.153 & 0.384 & -0.047 & 0.477     \\ 
R$_2$ mean & 0.094 & -0.088 &  0.243 & 0.464\\
R$_1$ std & 0.275 & 0.361 & -0.051 & 0.235 \\
R$_2$ std & 0.269 & -0.109 & 0.049 & 0.486\\
R$_1$ CV & 0.417 & 0.219 & 0.387 & -0.162\\
R$_2$ CV & 0.468 & -0.093 & 0.070 & 0.603\\
R$_1$ duration & 0.306 & 0.111   & 0.417 & -0.296\\
R$_2$ duration & 0.692 & -0.411  & -0.206 & -0.098\\
R$_1$ area & 0.191 & 0.438 & 0.020 & 0.372\\
R$_2$ area & 0.326 & -0.212 & 0.094 & 0.288\\ \hline
\end{tabular}
\end{center}
\caption{Table of Eigenvalues}
\label{tab_Structure_Matrix}
\end{minipage}
\hfil
\begin{minipage}[b]{.45\textwidth}
\begin{center}
\begin{tabular}{lcccc} \hline
& \multicolumn{4}{c}{Function} \\ \cline{2-5}
& 1 & 2 & 3 & 4 \\ \cline{1-5}
R$_1$ mean 	& 0.260 	& 0.863 	& 0.611 	& 0.084  \\ 
R$_2$ mean 	& -0.837 	& -1.044	&  0.843 	& 0.460  \\
R$_1$ std 	& -1.033 	& -1.023 	& -4.612 	& -0.471  \\
R$_2$ std 	& 0.999 	& 1.046		& -1.165 	& -0.934  \\
R$_1$ CV 	& 17.142 	& 17.026 	& 32.236 	& -12.682 \\
R$_2$ CV 	& 3.964 	& -6.788	& 2.522 	& 12.729  \\
R$_1$ duration 	& 0.298 	& 0.225 	& 0.514 	& -0.358 \\
R$_2$ duration 	& 0.306 	& -0.203	& -0.140 	& -0.146 \\
R$_1$ area 	& 0.011 	& 0.017 	& -0.001 	& 0.018  \\
R$_2$ area 	& 0.002 	& -0.003	& 0.005 	& 0.012  \\
(Constant) 	& -10.805	& -2.414	& -10.974 	& 2.054\\ \hline
\end{tabular}
\end{center}
\caption{Canonical Discriminant Function Coefficient}
\label{tab_Discriminant_Function_coefficient}
\end{minipage}
\end{table}

\begin{figure}[hbt]
\centering
\includegraphics[width=0.8\columnwidth,trim=0mm 5mm 0mm 0mm]{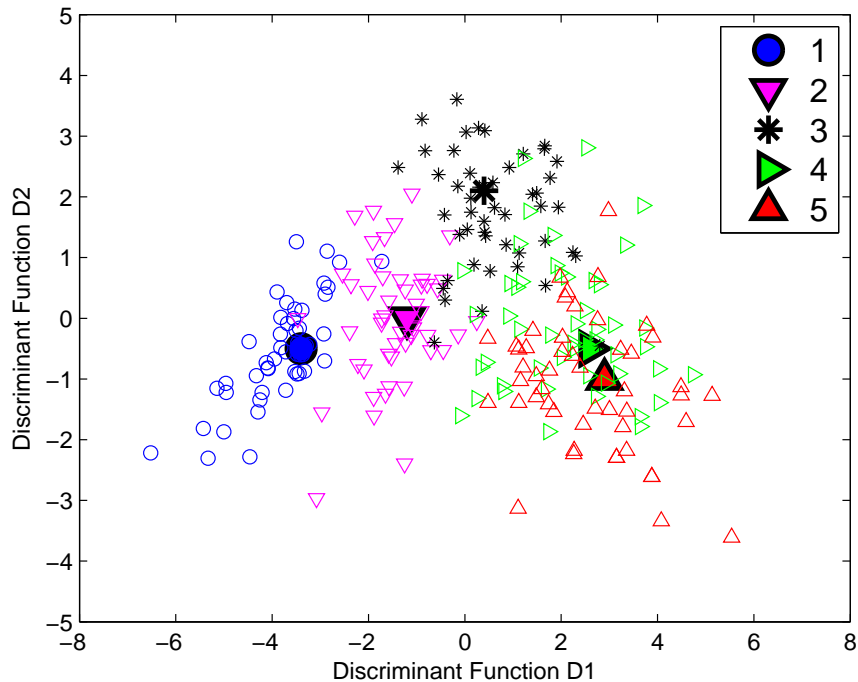}
\caption{Combined Group Plots of Discriminant Functions 1 and 2}
\label{fig_Combined_Group}
\end{figure}

For example, the discriminant function $D_1$ and $D_2$ are shown as follows:
\begin{equation}
\begin{split}
D_1  = & \left(0.26 \times R_1\,mean) + (-0.837 \times R_2\,mean) + (-1.033 \times R_1\,std) + (0.999 \times R_2\,std) \right. \\
	& \left.+\,(17.142 \times R_1\,CV) + (3.964 \times R_2\,CV) + (0.298 \times R_1\,duration) \right.\\
	& \left.+\,(0.306 \times R_2\,duration) + (0.011 \times R_1\,area) + (0.002 \times R_2\,area) - 10.805 \right.
\end{split}
\end{equation}
%
\begin{equation}
\begin{split}
D_2 = & \left(0.863 \times R_1\,mean) + (-1.044 \times R_2\,mean) + (-1.023 \times R_1\,std) + (1.046 \times R_2\,std) \!\!\right. \\
	& \left.+\,(17.026 \times R_1\,CV) + (-6.788 \times R_2\,CV) + (0.225 \times R_1\,duration) \right.\\
	& \left.+\,(-0.203 \times R_2\,duration) + (0.017 \times R_1\,area) + (-0.003 \times R_2\,area) - 2.414 \right.
\end{split}
\end{equation}
Fig.~\ref{fig_Combined_Group} plots all samples of discriminant `Function 1' and `Function 2' which cover 96.7\% of variance. The group centroids are plotted from left to right as the number of people increase. This is a good indication that the groups are well discriminated by functions $D_1$ and $D_2$.

Finally, the classification is performed based on the discriminant score. The classification results are shown in Table~\ref{tab_Classification_Results}. All correct prediction cases lie on the diagonal. The classification results show 81.6\% overall accuracy in detecting the number of people comprising a given group. Further, an overall accuracy of 97.9\% was achieved in predicting individual head counts. For example, the actual head count of 250 samples is 750 and the predicted head count was 734. 
\begin{table}[hbt]
\begin{center}
\begin{tabular}{lcccccll} \hline
	& NPeople	& \multicolumn{5}{c}{Predicted Group Membership} \\ \cline{3-8}
	& 	& 1 	& 2 	& 3 	& 4 	& 5 	& Total\\ \cline{1-8}
Count 	& 1 	& \textbf{47} 	& 3		& 0 	& 0  	& 0 	& 50\\ 
		& 2 	& 3 	& \textbf{46}	& 1 	& 0  	& 0		& 50\\ 
		& 3 	& 0 	& 3		& \textbf{44} 	& 2  	& 1		& 50\\ 
		& 4 	& 0 	& 3		& 7 	& \textbf{31}  	& 9		& 50\\ 
		& 5 	& 0 	& 0		& 0 	& 14  	& \textbf{36}	& 50\\ 
\cline{1-8}
\% 	& 1 	& \textbf{94} 	& 6		& 0 	& 0  	& 0 	& 100\\ 
	& 2 	& 6 	& \textbf{92}	& 2 	& 0  	& 0		& 100\\ 
	& 3 	& 0 	& 6		& \textbf{88} 	& 4  	& 2		& 100\\ 
	& 4 	& 0 	& 12	& 14 	& \textbf{62}  	& 18	& 100\\ 
	& 5 	& 0 	& 0		& 0 	& 28  	& \textbf{72}	& 100\\  \hline
\end{tabular}
\end{center}
\caption{Cassification Results}
\label{tab_Classification_Results}
\end{table}

\section{Multiple Object Detection under Uncontrolled Environment}
\begin{figure}[pbt]
\centering
\subfloat[Layout of Sensing Zone]{
\includegraphics[width=0.6\columnwidth,trim=0mm 10mm 0mm 0mm,clip]{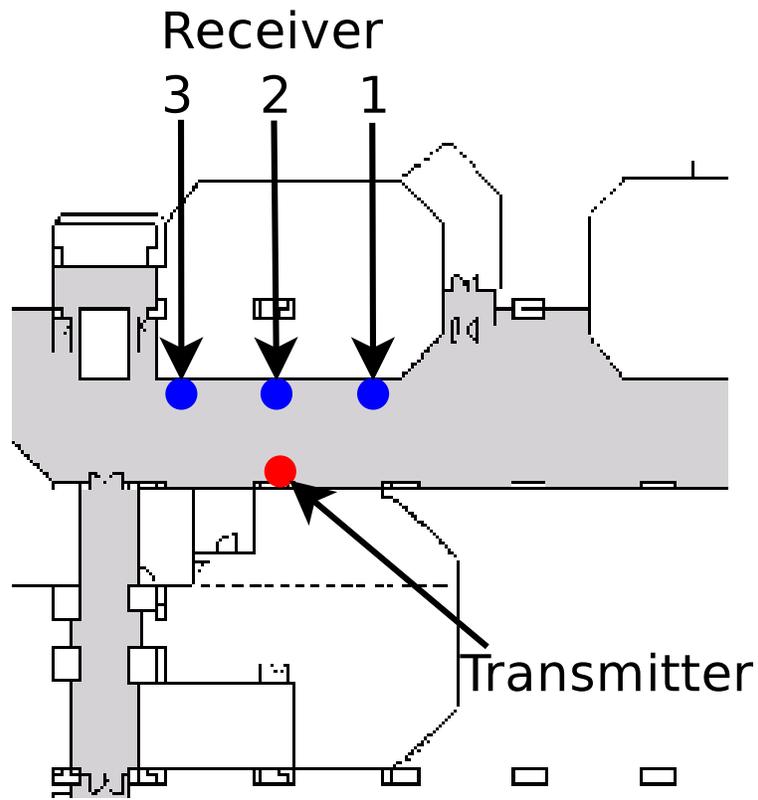}
\label{fig_Layout_Deployment_Site}}
\\
\subfloat[Experiment in progress with passerbys crossing]{
\includegraphics[width=0.78\columnwidth,trim=0mm 0mm 0mm 3mm]{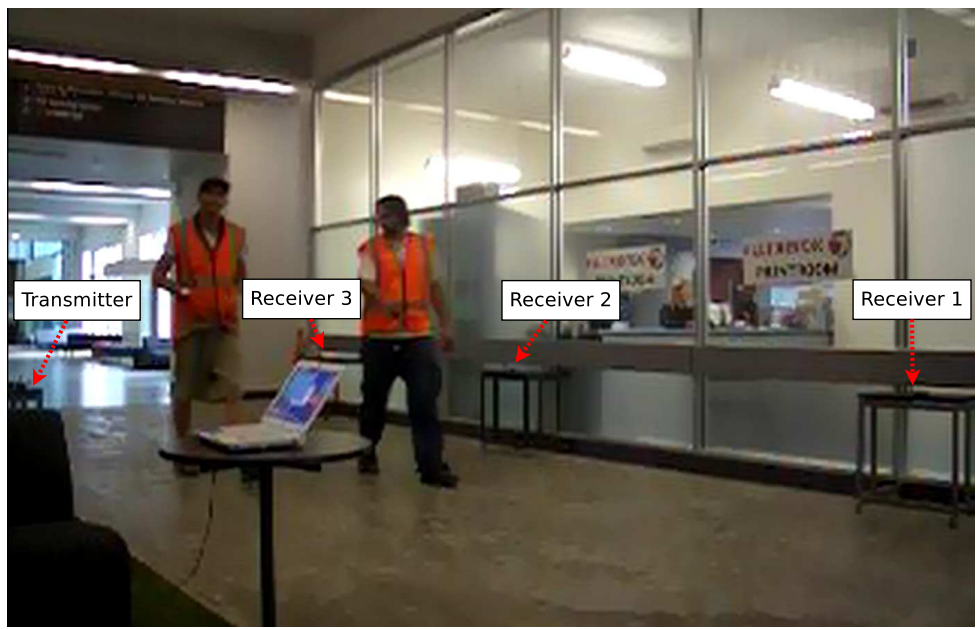}
\label{fig_Photo_Deployment_Site}}
\caption{Test Environment}
\label{fig_Sensor_Placement}
\end{figure}

In order to evaluate the accuracy in counting the number of people with no specific formation using the discriminant function that is calculated in previous section, we deployed the wireless sensor nodes in an uncontrolled environment. We deployed a configuration of one transmitter and three receivers, as shown in Fig.~\ref{fig_Layout_Deployment_Site} where the darker areas represent corridors and walkways. Three receivers were placed on one side of a wide corridor with one transmitter on the opposite side. The distance of T-R2 is 3m which allows up to five people walking past the corridor side-by-side. The distance of R1-R2 is 2.5m is the same as the distance of R2-R3. This sensor placement configuration can be considered as two sets of one-transmitter two-receiver configuration in symmetry. Each experiment was conducted for 15 minutes. The total number of people walking past was recorded on video to compare for verification of the detection result. A video frame of an experiment in progress is shown in Fig.\ref{fig_Photo_Deployment_Site}.

\begin{figure}[hbt]
\centering
\includegraphics[width=0.8\columnwidth,trim=0mm 10mm 0mm 0mm]{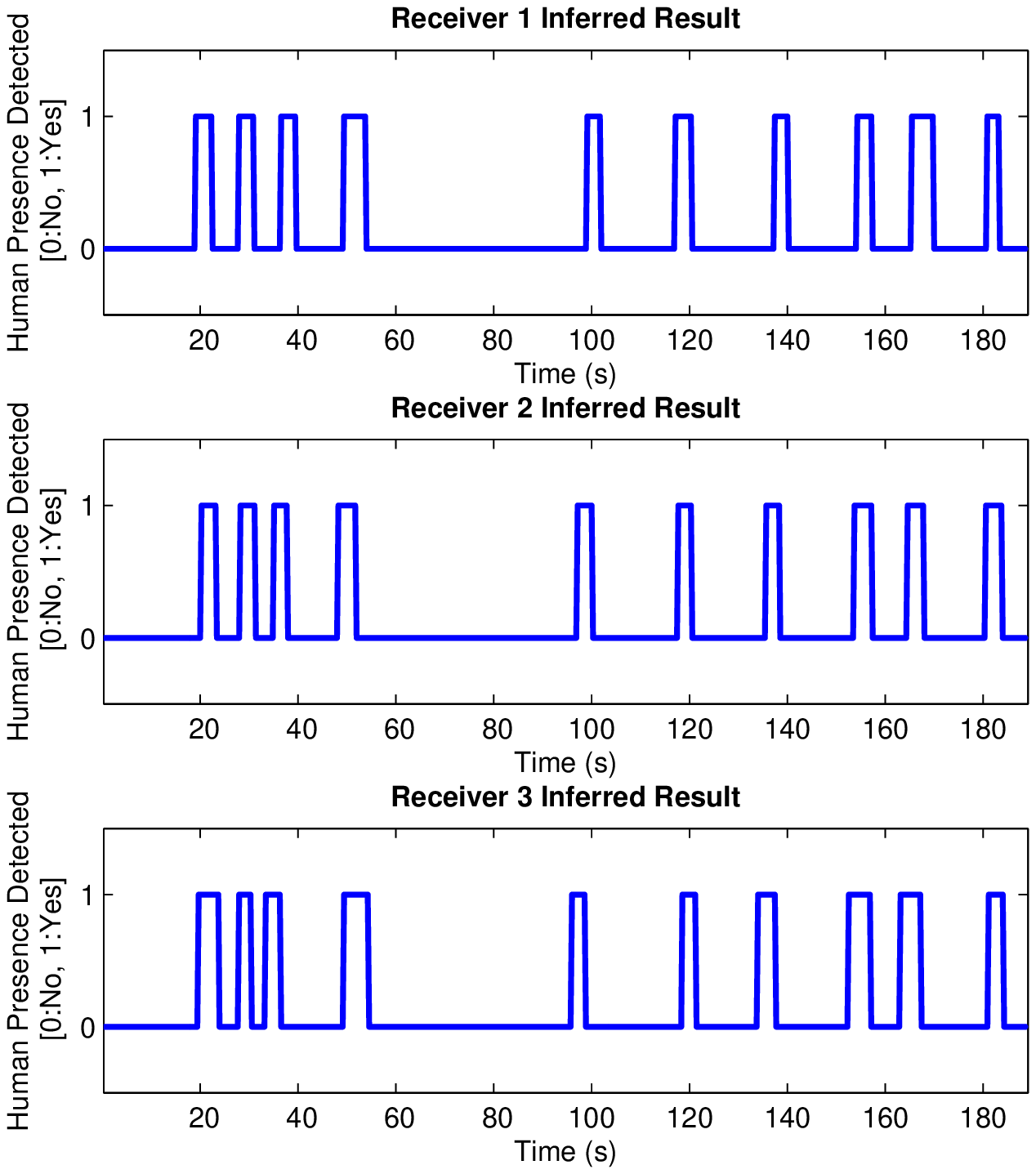}
\caption{Inferring Results under Uncontrolled Environment}
\label{fig_Uncontrolled_Environment_Result}
\end{figure}

For convenience in data analysis and comparison, the data collection was separated into 3-minute periods. Fig.~\ref{fig_Uncontrolled_Environment_Result} shows the results for one data collection period, during which our proposed detection system has been able to correctly infer every single occurrence of passerbys crossing the sensing zone. The recorded video shows the sequence of people in groups walking past the corridor as \{2, 1, 1, 3, 1, 1, 1, 1, 3, 1\}. The detection duration difference between receivers increases for 1st, 4th and 9th positive detections. As discussed previously, this is caused by the larger interference zone when more than one person passes, as shown in Fig.~\ref{fig_One-Txr-Two-Rxr}. 

\begin{figure}[hbt]
\centering
\includegraphics[width=0.8\columnwidth,trim=0mm 5mm 0mm 5mm]{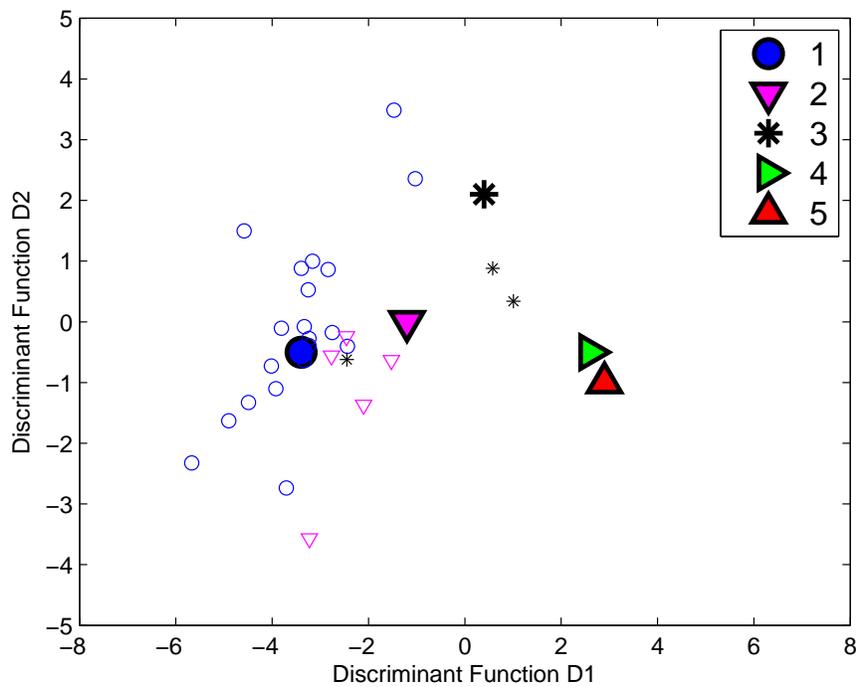}
\caption{Combined Group Plots of Discriminant Functions 1 \& 2 for R1-R2}
\label{fig_Combined_Group_Uncontrolled_R1R2}
\end{figure}

\begin{figure}[H]
\centering
\includegraphics[width=0.8\columnwidth,trim=0mm 5mm 0mm 5mm]{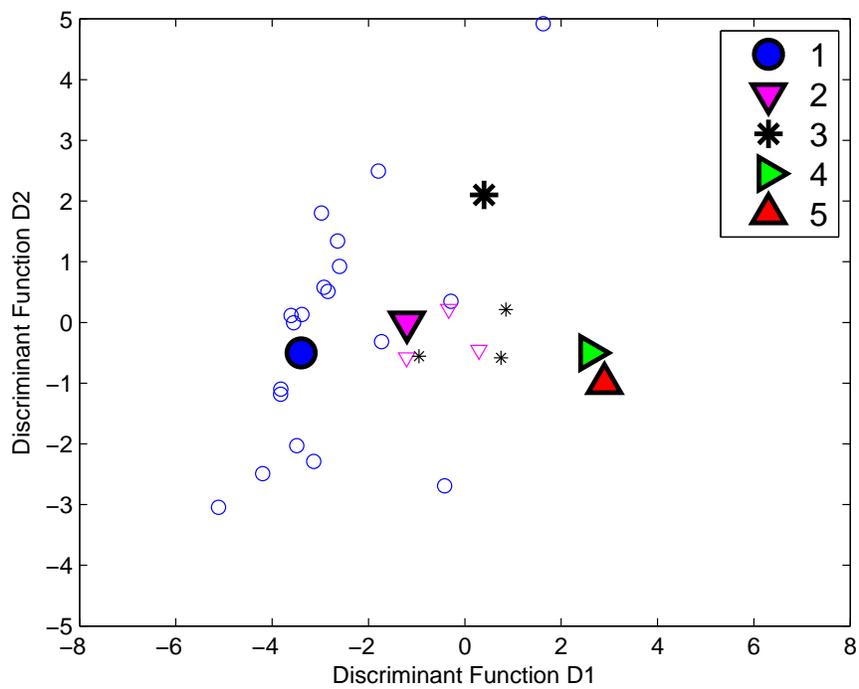}
\caption{Combined Group Plots of Discriminant Functions 1 \& 2 for R3-R2}
\label{fig_Combined_Group_Uncontrolled_R2R3}
\end{figure}

Using the one-transmitter three-receiver configuration, two sets of statistical information of positive detections can be extracted since it contains two sets of one-transmitter two-receiver configuration. When applying new cases, the discriminant score will be calculated using the discriminant functions that are constructed during training phase. All new samples' discriminant scores of discriminant `Function 1' and `Function 2' for R1-R2 and R3-R2 are plotted in Fig.~\ref{fig_Combined_Group_Uncontrolled_R1R2} and Fig.~\ref{fig_Combined_Group_Uncontrolled_R2R3} respectively. A sample can be classified when its discriminant score lies on the diagonal to the group centroid which is shown as a filled marker. 

We select a sample of the experiments for discussion, where a total of 39 people walked past the sensing zone in groups ranging from one to three persons; as this was an uncontrolled environment, we had not been able to detect groups of larger sizes although some of the instances could have been classified as a larger group.


First, we present the classification results for R1-R2 which are summarized in Table~\ref{tab_Classification_Results_Uncontrolled_R1R2} . From the R1-R2 dataset, the accuracy of predicting the number of people comprising a given group is 78.6\%, which is slightly lower than that for the controlled environment. This is expected as a few people walking in close proximity can be grouped in different ways from different angles and their relative positions with the group can change dynamically. More importantly, we aim to estimate the number of people in total, and that we have been able to achieve with an accuracy of around 97.4\% for this experiment, and with similar accuracies for other experiments. This is very close to the accuracy achieved in the controlled environment.

\begin{table}[hbt]
\begin{center}
\begin{tabular}{lcccccll}\hline
	& NPeople	& \multicolumn{5}{c}{Predicted Group Membership} \\ \cline{3-8}
		& 		& 1 	& 2 	& 3 	& 4 	& 5 	& Total\\ \cline{1-8}
Count 	& 1 	& \textbf{17} 	& 1		& 2 	& 0  	& 0 	& 20\\ 
		& 2 	& 2 	& \textbf{3}		& 0 	& 0  	& 0		& 5\\ 
		& 3 	& 1 	& 0		& \textbf{2} 	& 0  	& 0		& 3\\ 
		& 4 	& 0 	& 0		& 0 	& 0  	& 0		& 0\\ 
		& 5 	& 0 	& 0		& 0 	& 0  	& 0		& 0\\ 
\cline{1-8}
\% 		& 1 	& \textbf{85} 	& 5		& 10 	& 0  	& 0 	& 100\\ 
		& 2 	& 40 	& \textbf{60}	& 0 	& 0  	& 0		& 100\\ 
		& 3 	& 33.3 	& 0		& \textbf{66.7} 	& 0 	& 0		& 100\\ 
		& 4 	& 0 	& 0		& 0 	& 0  	& 0		& 0\\ 
		& 5 	& 0 	& 0		& 0 	& 0  	& 0		& 0\\ \hline
\end{tabular}
\end{center}
\caption{R1-R2 Prediction Results under Uncontrolled Environment}
\label{tab_Classification_Results_Uncontrolled_R1R2}
\end{table}

Next, we present the data from R3-R2, tabulated in Table~\ref{tab_Classification_Results_Uncontrolled_R2R3}. The accuracy of group size estimation dropped to 60.7\% for the same reasons as the R1-R2 case. The accuracy for number of people detected is 71.8\% with a noticeable degree of over-counting. Upon careful analysis of the video, we identified a possible cause of the over-counting that led to the higher estimation error. From Fig.~\ref{fig_Layout_Deployment_Site}, we note that R3 is located next to a stairwell on the left. That is a highly utilized staircase and, as a result, many people pass close to R3 but not necessarily continue right along the corridor across the R3-R2 sensing zone. This has resulted in two instances of over-estimating the group size by more than one, where a group of two people has been predicted as a group of four people. E.g., if a group four people come down the stairs and split into two groups of two, with one group continuing right along the corridor across the R3-R2 sensing zone while the other, moving in the opposite direction or straight across to the walkway on the lower left of Fig.~\ref{fig_Layout_Deployment_Site}, then there is a higher probability of the group size prediction to be affected. 

\begin{table}[t]
\begin{center}
\begin{tabular}{lcccccll}\hline
	& NPeople	& \multicolumn{5}{c}{Predicted Group Membership} \\ \cline{3-8}
		& 		& 1 	& 2 	& 3 	& 4 	& 5 	& Total\\ \cline{1-8}
Count 	& 1 	& \textbf{14} 	& 4		& 2 	& 0  	& 0 	& 20\\ 
		& 2 	& 1 	& \textbf{2}		& 0 	& 2  	& 0		&  5\\ 
		& 3 	& 0 	& 1		& \textbf{1} 	& 1  	& 0		& 3\\ 
		& 4 	& 0 	& 0		& 0 	& 0  	& 0		& 0\\ 
		& 5 	& 0 	& 0		& 0 	& 0  	& 0		& 0\\ 
\cline{1-8}
\% 		& 1 	& \textbf{70} 	& 20	& 10 	& 0  	& 0 	& 100\\ 
		& 2 	& 20 	& \textbf{40}	& 0 	& 40  	& 0		& 100\\ 
		& 3 	& 0 	& 33.3	& \textbf{33.3} 	& 33.3 	& 0		& 100\\ 
		& 4 	& 0 	& 0		& 0 	& 0  	& 0		& 0\\ 
		& 5 	& 0 	& 0		& 0 	& 0  	& 0		& 0\\ \hline
\end{tabular}
\end{center}
\caption{R3-R2 Prediction Results under Uncontrolled Environment}
\label{tab_Classification_Results_Uncontrolled_R2R3}
\end{table}

Based on the assumption that we intend to utilize pre-deployed wireless communication devices, the R3-R2 case highlights the need for careful selection of transmitter-receiver combinations in order to minimize prediction errors. For the R1-R2 case, although R1 is also near the building entrance/exit on the right, its position is sufficiently far away such that people moving in and out of the entrance/exit will not have a significant effect, if at all, on its prediction. 

\section{Conclusion}
The use of radio irregularity resulting from the movement of human objects crossing the path of a radio signal to detect human presence has been demonstrated previously and applied to intrusion detection~\cite{Lee:MST2010,Youssef:Mobicom2007,Lin:PIMRC2011}. 
However, the ability to detect more than one person remains a challenge if we rely on the characteristics of one signal's fluctuations. With pervasive networking brought about by large cyber-physical systems like the Internet of Things, the presence of numerous wireless communication devices allow us to study the fluctuations of multiple signals in close proximity of one another as a result of human interference and deduce the number of human objects that have crossed the paths of these signals.

In this paper, we have proposed a network-oriented approach that utilizes received signal strength information (RSSI) of received packets to detect and count people when they cross the signal transmission paths. This information can be easily obtained from device drivers of wireless network interfaces when the packets are received and the goal of our approach is to be able to easily utilize the existing wireless transmitters and receivers already deployed in the environment. Our approach which is based on the RSSI fluctuations between consecutive packets does not require accurate channel models nor complex signal processing techniques. It only needs to be trained to detect the RSSI fluctuation patterns associated with the objects of interest, e.g. groups of different number of people; no additional tuning is required during deployment.

Using a simple configuration of two receivers deployed in close proximity to each other, we have first demonstrated the ability to detect two persons walking side-by-side along a typical 1.5m wide corridor using a straightforward approach based on the difference in the periods of fluctuations experienced by the two signals paths as the two human subjects pass. We then extended our scheme to detect more human subjects using the same two-receiver configuration 
together with discriminant analysis to process the signal fluctuation data, we validated our scheme in a controlled environment and showed that it is able to accurately detect and count up to five persons with an accuracy of almost 98\%. \textcolor{black}{Next, we deployed our scheme in a public area without the ability to control the mobility patterns and group structure of passerbys, and achieved comparable accuracy in counting people. However, we also note that a poorly located receiver can induce high estimation errors and significantly reduce the accuracy of the system.} Ideally, a robust placement strategy should be chosen~\cite{Krause:TOSN2011}. However, in our target scenario, achieving the best device placements from among the already deployed wireless devices may not always be possible. Additional wireless devices need to be deployed to complement the existing topology, if the critical voids in detection need to be covered.

From this study, we aim to show that a large cyber-physical system, like the Internet of Things, can be exploited for applications like people counting without the need for specialized hardware. However, our method is not aimed to completely replace the specialized hardware for automated people counting but more as a complementary technology. While the scheme in its current form requires further work to enhance its capabilities further, it presents an exciting opportunity to turn an existing wireless communications network into a sensing system for automated people counting. 

As our ongoing and future work, we are extending the scheme for automated people counting in outdoor environments, e.g. to count visitors in public parks~\cite{Puccinelli:PerSens2011}, crowd size estimation~\cite{Hou:ICAL2008}, etc. Such technologies are increasingly being deployed for crowd size estimation to assist in crowd control and prevent any potential problems arising from loss of control over crowd size. Often, agencies involved in crowd safety and management require quick estimates to assist personnel on the ground, but most of the available technology rely on image and video processing which are complex and expensive.

\ifCLASSOPTIONcaptionsoff
  \newpage
\fi



%
\bibliographystyle{IEEEtran}
\bibliography{wmd}

%


%
%
%




\end{document}